\documentclass[iop]{emulateapj}
\usepackage{color}
\usepackage{natbib}
\usepackage{ulem}

\slugcomment{Draft version July 1}

\shorttitle{UV-NIR transmission spectrum of Earth as an exoplanet}
\shortauthors{B\'{e}tr\'{e}mieux \& Kaltenegger}

\begin{document}


\title{Transmission spectrum of Earth 
as a transiting exoplanet\\ from the ultraviolet to the near-infrared}


\author{Y. B\'{e}tr\'{e}mieux\altaffilmark{1} and L. Kaltenegger\altaffilmark{1,2}}
\altaffiltext{1}{MPIA, Koenigstuhl 17, Heidelberg, D 69117, Germany}
\altaffiltext{2}{Harvard-Smithsonian Center for Astrophysics, MS-20, 60 Garden Street, Cambridge, MA 02138, USA}
\email{betremieux@mpia.de, kaltenegger@mpia.de}



\begin{abstract}

Transmission spectroscopy of exoplanets is a tool to characterize rocky planets 
and explore their habitability. Using the Earth itself as a proxy, we model the atmospheric
cross section as a function of wavelength, and show the effect 
of each atmospheric species, Rayleigh scattering and refraction from 115 to 1000~nm. 
Clouds do not significantly affect this picture because refraction prevents the lowest 12.75~km 
of the atmosphere, in a transiting geometry for an Earth-Sun analog, to be sampled by a 
distant observer.
We calculate the effective planetary radius for the primary eclipse spectrum of an 
Earth-like exoplanet around a Sun-like star. Below 200 nm, ultraviolet~(UV) O$_2$ absorption 
increases the effective planetary radius by about 180~km, versus
27 km at 760.3 nm, and 14 km in the near-infrared (NIR) due predominantly to refraction. This translates into a 2.6\% change in effective planetary radius 
over the UV-NIR wavelength range, showing that the ultraviolet is an interesting wavelength range 
for future space missions.

\end{abstract}





\section{Introduction}

Many planets smaller than Earth have now been detected with the Kepler mission, and with the realization 
that small planets are much more numerous than giant ones (Batalha et al. 2013), 
future space missions, such as the James Webb Space Telescope (JWST), are being planned to characterize 
the atmosphere of potential Earth analogs by transiting spectroscopy, explore their  
habitability, and search for signs of life. The simultaneous detection of large abundances of 
either O$_{2}$ or O$_{3}$ in conjunction with a reducing species such as CH$_{4}$, or N$_2$O, 
are biosignatures on Earth (see e.g. Des Marais et al. 2002; Kaltenegger et al. 2010a and reference 
therein). Although not a clear indicative for the presence of life, H$_2$O is essential for life.

Simulations of the Earth's spectrum as a transiting exoplanet (Ehrenreich et al. 2006; 
Kaltenegger \& Traub 2009; Pall\'{e} et al. 2009; Vidal-Madjar et al. 2010; Rauer et al. 2011; 
Garc\'ia Mu\~{n}oz et al. 2012; Hedelt et al. 2013) have focused primarily on the visible (VIS) to the 
infrared (IR), the wavelength range of JWST (600-5000~nm). No models of spectroscopic 
signatures of a transiting Earth have yet been computed from the mid- (MUV) to the far-ultraviolet (FUV).
Which molecular signatures dominate this spectral range? In this paper, we present a model of a transiting 
Earth's transmission spectrum from 115 to 1000~nm (UV-NIR) during primary eclipse.
While no UV missions are currently in preparation, this model can serve as a basis for 
future UV mission concept studies.

\section{Model description} \label{model}

To simulate the spectroscopic signatures of an Earth-analog transiting its star, we 
modified the Smithsonian Astrophysical Observatory 1998 (SAO98) radiative transfer code 
(see Traub \& Stier 1976; Johnson et al. 1995; Traub \& Jucks 2002; Kaltenegger \& Traub 2009 
and references therein for details), which computes the atmospheric transmission of stellar radiation 
at high spectral resolution from a molecular line list database. Updates include a new database of 
continuous absorber's cross sections, as well as N$_2$, O$_2$, Ar, and CO$_2$ Rayleigh scattering 
cross sections from the ultraviolet (UV) to the near-infrared (NIR). A new module interpolates 
these cross sections and derives resulting optical depths according to the mole
fraction of the continuous absorbers and the Rayleigh scatterers in each atmospheric layer. 
We also compute the deflection of rays by atmospheric refraction to exclude 
atmospheric regions for which no rays from the star can reach the observer due to the observing 
geometry.

Our database of continuous absorbers is based on
the MPI-Mainz-UV-VIS Spectral Atlas of Gaseous Molecules\footnote{Hannelore Keller-Rudek, Geert K. Moortgat,
MPI-Mainz-UV-VIS Spectral Atlas of Gaseous Molecules, www.atmosphere.mpg.de/spectral-atlas-mainz}. 
For each molecular species of interest (O$_2$, O$_3$, CO$_2$, CO, CH$_4$, H$_2$O, NO$_2$, N$_2$O, and SO$_2$), 
we created model cross sections composed of several measured cross sections from different spectral 
regions, at different temperatures when measurements are available, with priority given to higher spectral resolution
measurements (see Table~\ref{tbl_crsc}). We compute absorption optical depths for different altitudes in the atmosphere 
using the cross section model with the closest temperature to that of the atmospheric layer considered. 
Note that we do not consider line absorption from atomic or ionic species which could 
produce very narrow but possibly detectable features at high spectral resolution (see also Snellen et al. 2013).

The Rayleigh cross sections, $\sigma_R$, of N$_2$, O$_2$, Ar, and CO$_2$, 
which make-up 99.999\% of the Earth's atmosphere, are computed with
\begin{equation} \label{rayl}
{\sigma_{R}} = \frac{32\pi^{3}}{3} \left( \frac{{\nu_{0}}}{n_{0}} \right)^{2} w^{4} F_K ,
\end{equation}
where $\nu_0$ is the refractivity at standard pressure and temperature (or standard refractivity) of the 
molecular species, $w$ is the wavenumber, $F_K$ is the King correction factor, and $n_{0}$ is Loschmidt's constant.
Various parametrized functions are used to describe the spectral dependence of $\nu_0$ and $F_K$.
Table~\ref{tbl_rayl} gives references for the functional form of both parameters, as well as their 
spectral region.

The transmission of each atmospheric layer is computed with Beer's law from all optical depths. 
We use disc-averaged quantities for our model atmosphere. 

We use a present-day Earth vertical composition 
(Kaltenegger et al. (2010b) for SO$_2$; Lodders \& Fegley, Jr. (1998) for Ar; and Cox (2000) for all other 
molecules) up to 130~km altitude, unless specified otherwise. Above 130~km, we assume constant mole fraction 
with height for all the molecules except for SO$_2$ which we fix at zero, and for N$_2$, O$_2$, and Ar which
are described below. Below 100~km, we use the US 1976 atmosphere (COESA 1976) as the temperature-pressure 
profile. Above 100~km, the atmospheric density is sensitive to and increases with solar activity (Hedin 1987).
We use the tabulated results of the MSIS-86 model, for solar maximum (Table A1.2) and solar minimum conditions (Table A1.1) 
published in Rees~(1989), to derive the atmospheric density, pressure, and mole fractions for 
N$_2$, O$_2$, and Ar above 100~km. We run our simulations in two different spectral regimes. In the VIS-NIR, 
from 10000 to 25000~cm$^{-1}$ (400-1000~nm), we use a 0.05~cm$^{-1}$ grid, while in the UV from 
25000 to 90000~cm$^{-1}$ (111-400~nm), we use a 0.5~cm$^{-1}$ grid. For displaying the results, the 
VIS-NIR and the UV simulations are binned on a 4~cm$^{-1}$ and a 20~cm$^{-1}$ grid, respectively. 
The choice in spectral resolution impacts predominantly the detectability of spectral features.

The column abundance of each species along a given ray is computed taking into account refraction, 
tracing specified rays from the observer back 
to their source. Each ray intersects the top of the model atmosphere with an impact parameter $b$, 
the projected radial distance of the ray to the center of the planetary disc as viewed by the observer. 
As rays travel through the planetary atmosphere, they are bent by refraction
along paths define by an invariant $L = (1 + \nu(r)) r \sin\theta(r)$  where both the zenith angle, 
$\theta(r)$, of the ray, and the refractivity, $\nu(r)$, are functions of the radial position of the 
ray with respect to the center of the planet. The refractivity is given by 
\begin{equation} \label{refrac}
\nu(r) = \left( \frac{n(r)}{n_{0}} \right) \sum_{j} f_{j}(r) {\nu_{0}}_{j} = \left( \frac{n(r)}{n_{0}} \right) \nu_{0}(r) ,
\end{equation}
where ${\nu_{0}}_{j}$ is the standard refractivity of the j$^{th}$ molecular species while $\nu_{0}(r)$ is that of the 
atmosphere, $n(r)$ is the local number density, and $f_{j}(r)$ is the mole fraction of the j$^{th}$ species.
Here, we only consider the main contributor to the refractivity (N$_2$, O$_2$, Ar, and CO$_2$) which are well-mixed 
in the Earth's atmosphere, and fix the standard refractivity at all altitudes at its surface value. 

If we assume a zero refractivity at the top of the atmosphere, the minimum radial position from the 
planet's center, $r_{min}$, that can be reached by a ray is related to its impact parameter by
\begin{equation} \label{refpath}
L = (1 + \nu(r_{min})) r_{min} = R_{top} \sin\theta_{0} = b ,
\end{equation}
where $R_{top}$ is the radial position of the top of the atmosphere and $\theta_{0}$ is the 
zenith angle of the ray at the top of the atmosphere. Note that $b$ is always larger than $r_{min}$, 
therefore the planet appears slightly larger to a distant observer. For each ray, we specify $r_{min}$, 
compute $\nu(r_{min})$, and obtain the corresponding impact parameter. Then, each ray is traced
through the atmosphere every 0.1~km altitude increment, and column abundances, average mole fractions, 
as well as cumulative deflection along the ray are computed for each atmospheric 
layer (Johnson et al. 1995; Kaltenegger \& Traub 2009). 

We characterize the transmission spectrum of the exoplanet using effective atmospheric thickness, 
$\Delta z_{eff}$, the increase in planetary radius due to atmospheric absorption during primary eclipse. 
To compute $\Delta z_{eff}$ for an exoplanet, we first specify $r_{min}$ for $N$ rays spaced in constant 
altitude increments over the atmospheric region of interest. We then compute the transmission, $T$, and 
impact parameter, $b$, of each ray through the atmosphere, and finally use,
\begin{mathletters}
\begin{eqnarray}
R_{eff}^{2} = R_{top}^{2} - \sum_{i = 1}^{N} \left( \frac{T_{i+1} + T_{i}}{2} \right) (b_{i+1}^{2} - b_{i}^{2}) \label{reff} \\
R_{top} = R_{p} + \Delta z_{atm} \\
\Delta z_{eff} = R_{eff} - R_{p} , 
\end{eqnarray}
\end{mathletters}
where $R_{eff}$ is the effective radius of the planet, $R_{top}$ is the radial position of the top of the atmosphere, 
$R_{p}$ is the planetary radius (6371~km), $\Delta z_{atm}$ is the thickness of the atmosphere, and $i$ denotes 
the ray considered. Note that $(N+1)$ refer to a ray that grazes the top of the atmosphere. The rays define $N$ 
projected annuli whose transmission is the average of the values at the borders of the annulus. 
The top of the atmosphere is defined where the transmission is 1, and no bending occurs ($b_{N+1} = R_{top}$). 
We choose R$_{top}$ where atmospheric absorption and refraction are negligible, and use
100~km in the VIS-NIR, and 200~km in the UV for $\Delta z_{atm}$. 

To first order, the total deflection of a ray through an atmosphere is proportional 
to the refractivity of the deepest atmospheric layer reached by a ray (Goldsmith 1963). 
The planetary atmosphere density increases exponentially with depth, therefore some of the 
deeper atmospheric regions can bend all rays away from the observer (see e.g. Sidis \& Sari 2010, 
Garc\'ia Mu\~{n}oz et al. 2012), and will not be sampled by the observations. At which altitudes this occurs depends on 
the angular extent of the star with respect to the planet. For an Earth-Sun analog, rays that reach a distant observer 
are deflected on average no more than 0.269$\degr$. We calculate that 
the lowest altitude reached by these grazing rays range from about 14.62~km at 115~nm, 13.86~km at 198~nm (shortest 
wavelength for which all used molecular standard refractivities are measured), to 12.95~km at 400~nm, 
and 12.75~km at 1000~nm.

As this altitude is relatively constant in the VIS-NIR, we incorporate this effect in our model 
by excluding atmospheric layers below 12.75~km. 
To determine the effective planetary radius, we choose standard refractivities representative of the lowest 
opacities within each spectral region: 2.88$\times10^{-4}$ for the VIS-NIR, and 3.00$\times10^{-4}$ for the UV. 
We use 80 rays from 12.75 to 100~km in the VIS-NIR, and 80 rays from 12.75 to 200~km altitude in the UV. 
In the UV, the lowest atmospheric layers have a negligible transmission, thus the exact exclusion value of the 
lowest atmospheric layer, calculated to be between 14.62 and 12.75~km, do not impact the modeled UV spectrum.

\section{Results and discussion} \label{discussion}

The increase in planetary radius due to the additional atmospheric absorption of a transiting Earth-analog is 
shown in Fig.~\ref{spectrum} from 115 to 1000~nm. 

The individual contribution of Rayleigh scattering 
by N$_2$, O$_2$, Ar, and CO$_2$ is also shown, with and without the effect of refraction by these same species, 
respectively. The individual contribution of each species, shown both in the lower panel of Fig.~\ref{spectrum} and 
in Fig.~\ref{absorbers}, are computed by sampling all atmospheric layers down to the surface, assuming 
the species considered is the only one with a non-zero opacity. In the absence of absorption, the effective 
atmospheric thickness is about 1.8~km, rather than zero, because the bending of the rays due to refraction 
makes the planet appear larger to a distant observer. 

The spectral region shortward of 200~nm is shaped by O$_2$ absorption and depends on solar activity. 
Amongst the strongest O$_2$ features are two narrow peaks around 120.5, and 124.4~nm, which, increase the planetary radius by 
179-185 and 191-195~km, respectively. The strongest O$_2$ feature, the broad Schumann-Runge continuum, increases the planetary
radius by more than 150~km from 134.4 to 165.5~nm, and peaks around 177-183~km. 
The Schumann-Runge bands, from 180 to 200~nm, create maximum variations of 30~km in the effective planetary 
radius. O$_2$ features can also be seen in the VIS-NIR, but these are much 
smaller than in the UV. Two narrows peaks around 687.0 and 760.3~nm increase the planetary radius to about 
27~km, at the spectral resolution of the simulation. 

Ozone absorbs in two different broad spectral regions in the UV-NIR increasing the planetary 
radius by 66~km around 255~nm (Hartley band), and 31~km around 602~nm (Chappuis band). 
Narrow ozone absorption, from 310 to 360~nm (Huggins band), produce variations in the 
effective planetary radius no larger than 2.5~km. Weak ozone bands are also present throughout 
the VIS-NIR: all features not specifically identified on the small VIS-NIR panel in 
Fig.~\ref{spectrum} are O$_3$ features, and show changes in the 
effective planetary radius on the order of 1~km. 

NO$_2$ and H$_2$O are the only other molecular absorbers that create observable features in the spectrum
(Fig.~\ref{spectrum}, small VIS-NIR panel). NO$_2$ shows a very weak band system in the 
visible shortward of 510~nm, which produces less than 1~km variations in the effective planetary
radius. H$_2$O features are observable only around 940~nm, where they increase the effective planetary 
radius to about 14.5~km. 

Rayleigh scattering (Fig.~\ref{spectrum}) increases the planetary radius by about 68~km at 115~nm, 
27~km at 400~nm, and 5.5~km at 1000~nm, and dominates the spectrum from about
360 to 510~nm where few molecules in the Earth's atmosphere absorb, and refraction is not yet 
the dominant effect. In this spectral region, NO$_2$ is the dominant molecular absorber
but its absorption is much weaker than Rayleigh scattering. 

The lowest 12.75~km of the atmosphere is not accessible to a distant observer because no rays below that altitude 
can reach the observer in a transiting geometry for an Earth-Sun analog. Clouds located below that altitude 
do not influence the spectrum and can therefore be ignored in this geometry.
Figure~\ref{spectrum} also shows that refraction influences the observable spectrum for wavelengths larger than 400~nm.  

The combined effects of refraction and Rayleigh scattering increases the planetary radius by about 27~km at 400~nm,
16~km at 700~nm, and 14~km at 1000~nm.  In the UV, the lowest 12.75~km of the atmosphere 
have negligible transmission, so this atmospheric region can not be seen by a distant observer irrespective 
of refraction. 

Both Rayleigh scattering and refraction can mask some of the signatures from molecular species. 
For instance, the individual contribution of the H$_2$O band in the 900-1000~nm region can increase the 
planetary radius by about 10~km. However, H$_2$O is concentrated in the lowest 10-15~km of the Earth's atmosphere, 
the troposphere, hence its amount above 12.75~km increases the planetary radius above the refraction 
threshold only by about 1~km around 940~nm. 
The continuum around the visible O$_2$ features is due to the combined effects of Rayleigh scattering, ozone absorption, and refraction. It increases the effective planetary radius by about 21 and 17~km 
around 687.0 and 760.3~nm, respectively. The visible O$_2$ features add 6 and 10~km to the continuum 
values, at the spectral resolution of the simulation. 

{Figure~\ref{data} compares our effective model from Fig.~\ref{spectrum} with atmospheric thickness with the one
deduced by Vidal-Madjar et al. (2010) from Lunar eclipse data obtained in the penumbra. The contrast of the two O$_2$
features are comparable with those in the data. However, there is a slight offset (about 3.5~km) and a tilt 
in the main O$_3$ absorption profile. Note that, Vidal-Madjar et al. (2010) estimate that several sources of systematic errors and statistical uncertainties prevent them
from obtaining absolute values better than $\pm$2.5~km. Also, we do not include 
limb darkening in our calculations. However, for a transiting Earth, the
atmosphere eclipses an annular region on the Sun, whereas during Lunar eclipse
observations, it eclipses a band across the Sun (see Fig. 4 in Vidal-Madjar et al.
2010), leading to different limb darkening effects.}

Many molecules, such as CO$_2$, H$_2$O, CH$_4$, and CO, absorb ultraviolet radiation shortward 
of 200~nm (see Fig.~\ref{absorbers}). However, for Earth, the O$_2$ absorption dominates in this 
region and effectively masks their signatures. For planets without molecular oxygen, the far UV
would still show strong absorption features that increase the planet's effective radius by a 
higher percentage than in the VIS to NIR wavelength range.

\section{Conclusions}

\begin{figure*}
\epsscale{1.0}
\plotone{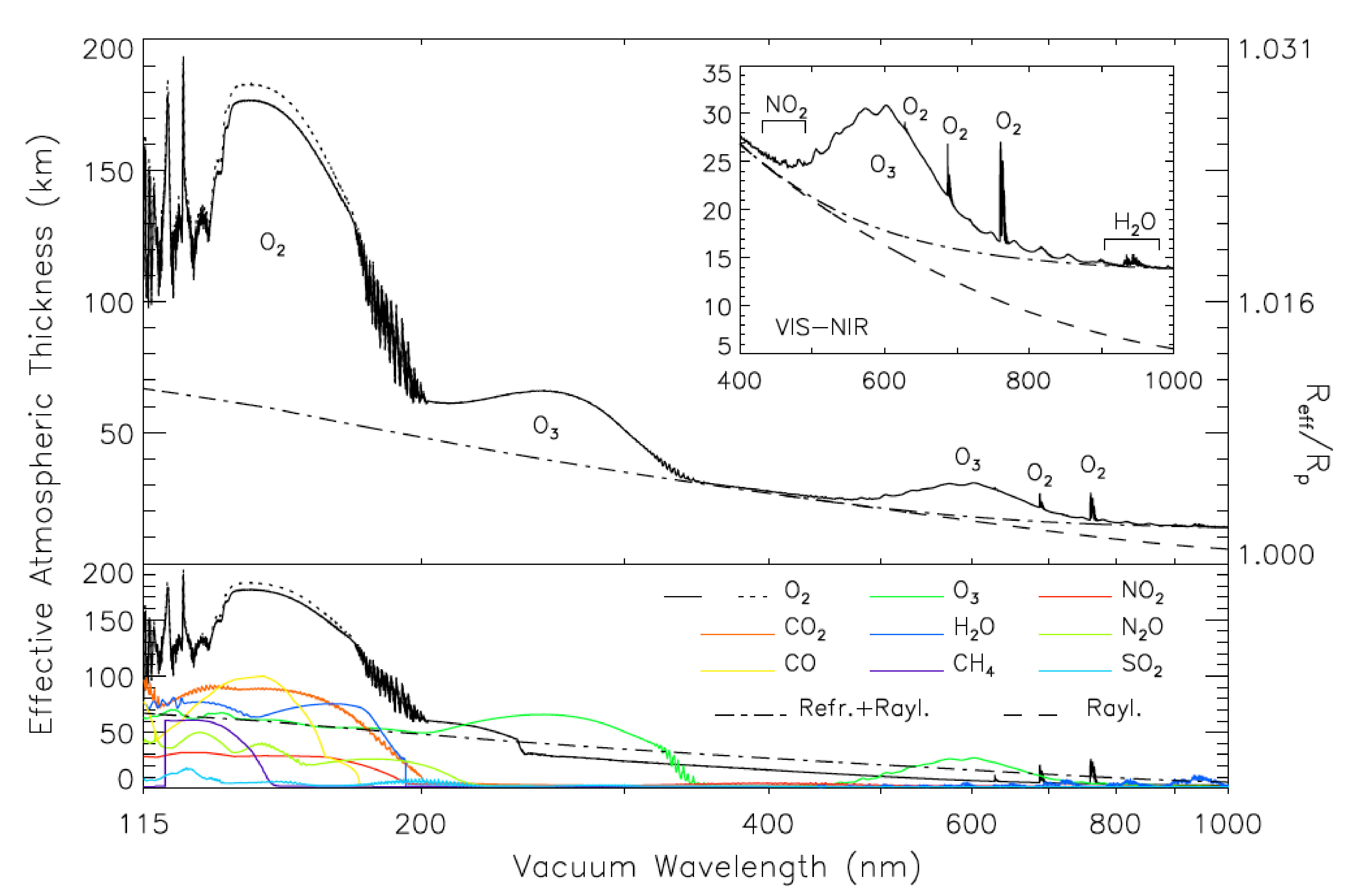}
\caption{Effective atmospheric thickness and effective planetary radius, R$_{eff}$
(expressed in Earth radii, R$_p$), of a transiting Earth
as a function of wavelength. The upper panel shows the overall spectrum from the 
UV to the NIR while the small insert panel zooms on the visible and NIR for clarity. Prominent 
spectral features are identified. The lower panel shows the effect of individual species (see Fig.~\ref{absorbers} for details).
In all panels, the solid and short-dashed lines are for solar minimum and solar maximum conditions, 
respectively. Furthermore, the dot-dashed and dashed lines 
show the individual contribution of Rayleigh scattering by N$_2$, O$_2$, Ar and CO$_2$, with and 
without the effect of refraction by these same species, respectively. See electronic edition of the Journal for a color version of this figure. \label{spectrum}}
\end{figure*}

\begin{figure*}
\epsscale{1.0}
\plotone{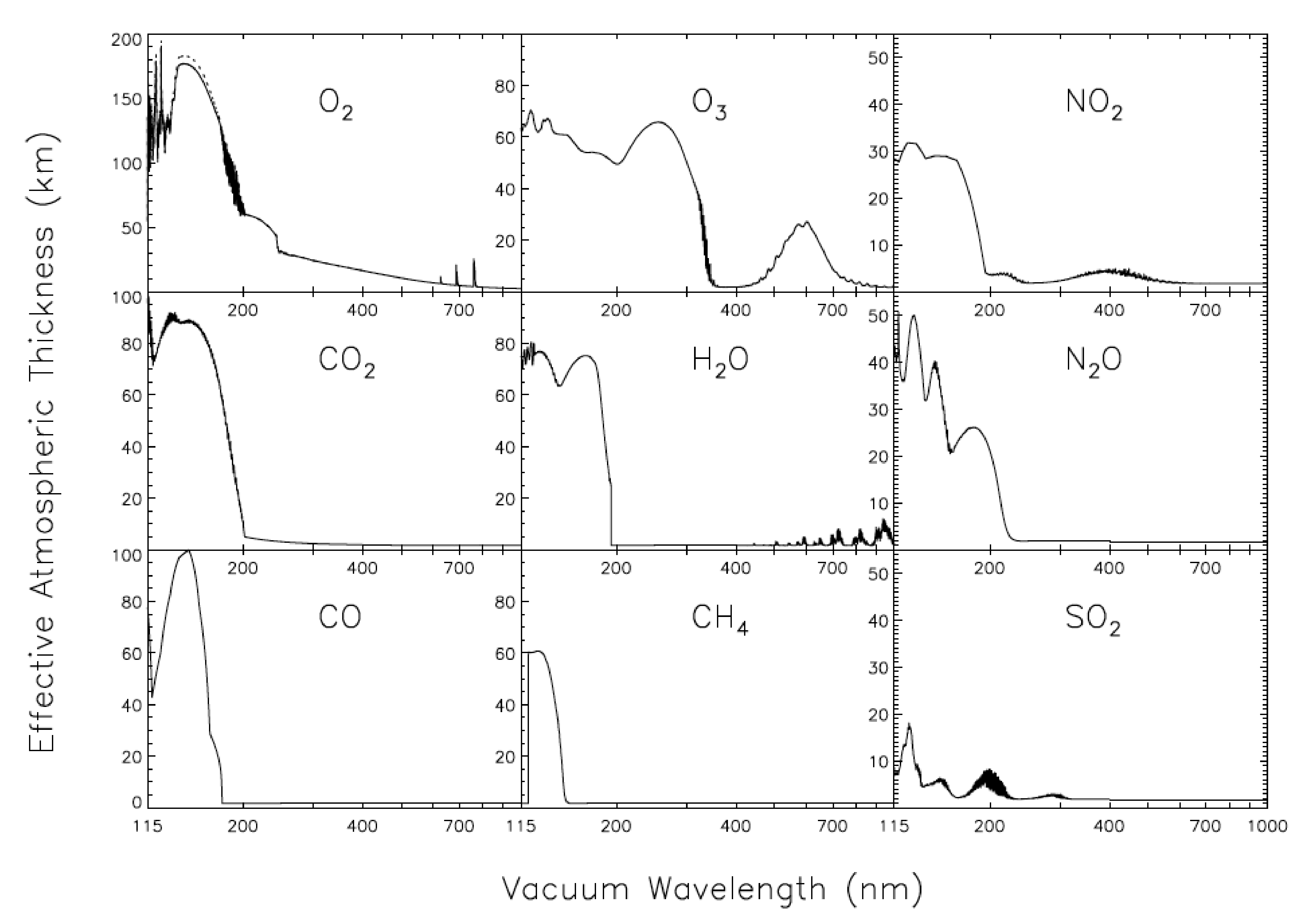}
\caption{Effective atmospheric thickness for each molecular species as a function of 
wavelength, assuming it is the only one with a non-zero opacity. Note the difference in scale on the 
y-axis: 200~km for O$_2$, 55~km for NO$_2$, N$_2$O, and SO$_2$, and 100~km elsewhere. 
O$_2$ and CO$_2$ curves also show the effect of Rayleigh scattering. The O$_2$ panel shows 
differences between solar minimum (solid line) and solar maximum (short-dashed line) 
conditions. \label{absorbers}}
\end{figure*}

\begin{figure*}
\epsscale{1.0}
\plotone{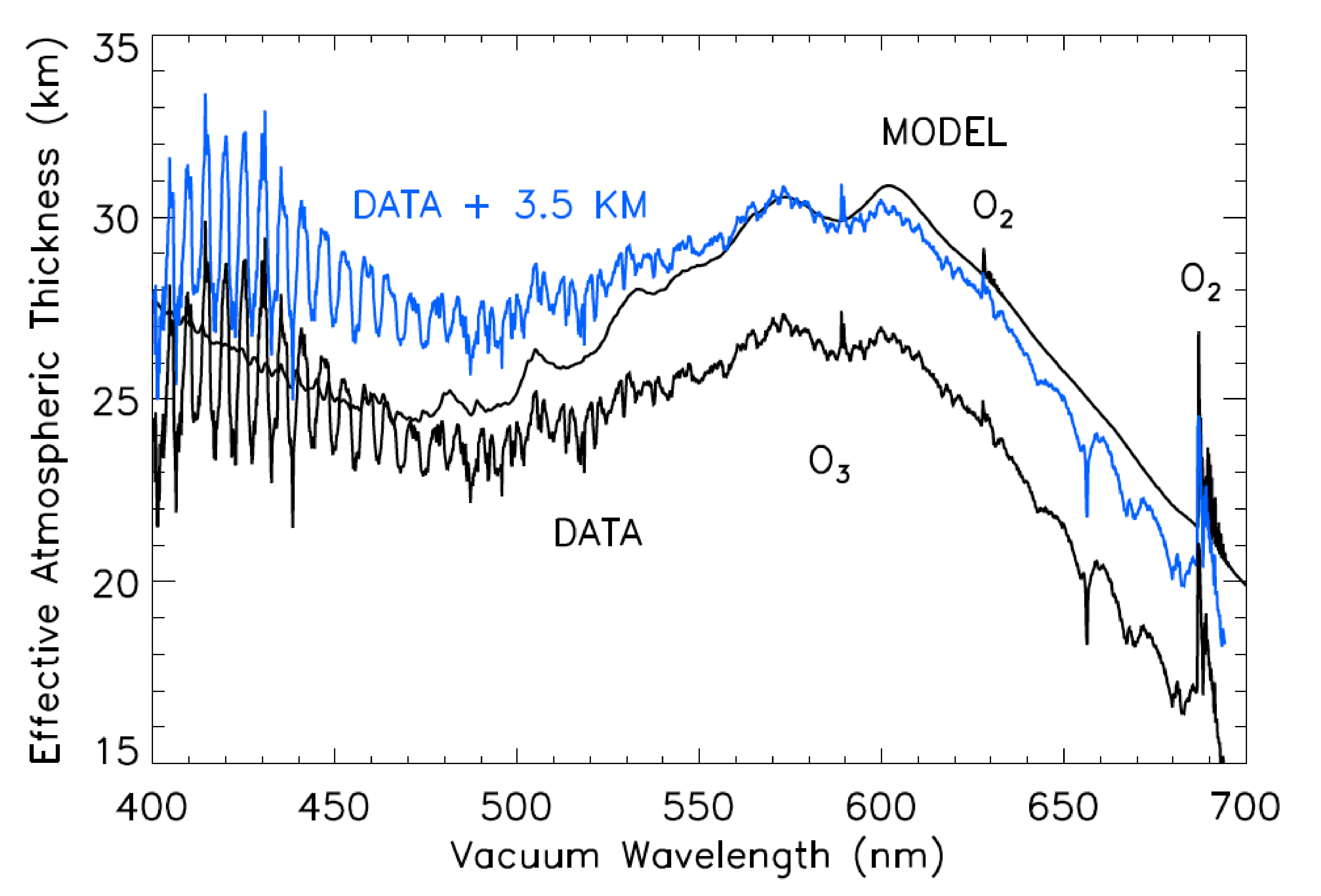}
\caption{Comparison of our model transiting Earth effective atmospheric thickness with the data of 
Vidal-Madjar et al. (2010). The data is also shown shifted by +3.5~km for ease of comparison. 
Note that the obvious oscillatory features on the short wavelength side of the data are due to uncorrected
echelle spectrograph cross-order contamination and should be ignored. \label{data}}
\end{figure*}

The UV-NIR spectrum (Fig.~\ref{spectrum}) of a transiting Earth-like exoplanet can be divided 
into 5 broad spectral regions characterized by the species or process that predominantly increase the 
planet's radius: one O$_2$ region (115-200~nm), two O$_3$ regions (200-360~nm and 510-700~nm), 
one Rayleigh scattering region (360-510~nm), and one refraction region (700-1000~nm). 

From 115 to 200~nm, O$_2$ absorption increases the effective planetary 
radius by up to 177-183~km, except for a narrow feature at 124.4~nm where it goes up to 191-195~km,
depending on solar conditions. Ozone increases the effective planetary radius up to 66~km in the 200-360~nm region, and 
up to 31~km in the 510-700~nm region. From 360 to 510~nm, Rayleigh scattering predominantly 
increases the effective planetary radius up to 31~km. Above 700~nm, refraction and Rayleigh 
scattering increase the effective planetary radius to a minimum of 14~km, masking H$_2$O
bands which only produce a further increase of at most 1~km. Narrow O$_2$ absorption bands 
around 687.0 and 760.3~nm, both increase the effective planetary radius by 27~km, that is 
6 and 10~km above the continuum, respectively. NO$_2$ only produces variations on the order 
of 1~km or less above the continuum between 400 and 510~nm.

One can use the NIR as a baseline against which the other regions in the UV-NIR 
can be compared to determine that an atmosphere exists. From the peak 
of the O$_2$ Schumann-Runge continuum in the FUV, to the NIR continuum, the 
effective planetary radius changes by about 166~km, which translates into a 2.6\% change. 

The increase in effective radius of the Earth in the UV due to O$_2$ absorption shows that this wavelength 
range is very interesting for future space missions. This increase in
effective planetary radius has to be traded off against the lower
available stellar flux in the UV as well as the instrument sensitivity at different wavelengths for
future mission studies. For habitable planets with atmospheres different 
from Earth's, other molecules, such as CO$_2$, H$_2$O, CH$_4$, and CO, would dominate the absorption 
in the ultraviolet radiation shortward of 200nm, providing an interesting alternative to 
explore a planet's atmosphere.



\acknowledgments{The authors acknowledge support from DFG funding ENP
  Ka 3142/1-1. The authors would also like to thank Jonathan McDowell
  for very useful suggestions and comments.}

\clearpage 

\begin{deluxetable}{cccc}
\tabletypesize{\scriptsize}
\tablecaption{UV - NIR model continuous absorption cross sections. \label{tbl_crsc}}
\tablewidth{0pt}
\tablehead{
\colhead{ Species } & \colhead{ Temperature of data (K) } 
& \colhead{ Wavelength range (nm)\tablenotemark{a}} & \colhead{ References }
}
\startdata
O$_{2}$  &       303          &  115.0 - 179.2    &    Lu et al. (2010) \\
    &       300          &  179.2 - 203.0    &    Yoshino et al. (1992) \\
    &       298          &  203.0 - 240.5    &    Yoshino et al. (1988) \\
    &       298          &  240.5 - 294.0    &    Fally et al. (2000) \\
O$_{3}$  &      298           &  110.4 - 150.0    &    Mason et al. (1996) \\
    &      298           &  150.0 - 194.0    &    Ackerman (1971) \\
    &      218           &  194.0 - 230.0    &   Brion et al. (1993) \\ 
    & 293, 273, 243, 223 &  230.0 - 1070.0   &   Bogumil et al. (2003) \\
NO$_{2}$ &    298             &   15.5 - 192.0    &    Au \& Brion (1997) \\
    &    298             &   192.0 - 200.0    &    Nakayama et al. (1959) \\
    &    298             &   200.0 - 219.0   &  Schneider et al. (1987) \\
    &    293             &   219.0 - 500.01\tablenotemark{b}  &   Jenouvrier et al. (1996) + M\'{e}rienne et al. (1995)\\
    & 293, 273, 243, 223 &   500.01\tablenotemark{b}   - 930.1\tablenotemark{c}     &    Bogumil et al. (2003) \\
CO  &  298               &  6.2 - 177.0      &    Chan et al. (1993) \\
CO$_{2}$ &   300              &  0.125 - 201.6    &     Huestis \& Berkowitz (2010) \\
H$_{2}$O &   298              &   114.8 - 193.9   &     Mota et al. (2005) \\
CH$_{4}$ &   295              &   120.0 - 142.5   &     Chen \& Wu (2004)     \\  
    &   295              &   142.5 - 152.0   &     Lee et al. (2001)    \\
N$_{2}$O & 298                &   108.2 - 172.5   &     Zelikoff et al. (1953) \\
    & 302, 263, 243, 225, 194 &   172.5 - 240.0\tablenotemark{d}   &     Selwyn et al. (1977)   \\
SO$_{2}$ & 293                &   106.1 - 171.95\tablenotemark{e}   &     Manatt \& Lane (1993) \\
    & 295\tablenotemark{g} &   171.95\tablenotemark{e} - 262.53\tablenotemark{f}   &     Wu et al. (2000)   \\
    & 358, 338, 318, 298\tablenotemark{h} &   262.53\tablenotemark{f} - 416.66   &  
Vandaele et al. (2009)\tablenotemark{i}   \\
\enddata
\tablenotetext{a}{Wavelength range determined by cross-over point between data sets, or wavelength coverage of data.}
\tablenotetext{b}{Value for the NO$_{2}$ 293~K model. Values are respectively 234.6, 234.0 and 234.2~nm for the~273, 
243 and 223~K models.}
\tablenotetext{c}{Value for the NO$_{2}$ 293~K model. For all other models, data is only available until 890.1~nm}
\tablenotetext{d}{Data is only available until 210.0~nm at 194~K.}
\tablenotetext{e}{Value is 175.55~nm for the 358~K SO$_{2}$ model.}
\tablenotetext{f}{Value for the 338 and 318~K models. 358, 298 and 200~K models use 
respectively 253.74, 262.45 and 297.55~nm.}
\tablenotetext{g}{Temperature of data used for the 338, 318, and 298~K SO$_{2}$ models. 
358 and 200~K models use respectively 400~K and 200~K data.
Note that the 400~K data has been shifted by -0.09~nm.}
\tablenotetext{h}{The 200~K SO$_{2}$ model uses the 203~K data from Bogumil et al. (2003) 
between 297.55 and 335.5~nm.}
\tablenotetext{i}{Data is shifted by +0.08~nm}

\end{deluxetable}

\clearpage 

\begin{deluxetable}{ccc}
\tabletypesize{\scriptsize}
\tablecaption{Parameters for Rayleigh cross section \label{tbl_rayl}}
\tablewidth{0pt}
\tablehead{\multicolumn{3}{c}{Standard refractivity} \\
\colhead{ Species } & \colhead{ Wavelength range (nm) }  & \colhead{ References } 
}
\startdata
N$_{2}$    &  149 - 189    &    Griesmann \& Burnett (1999) \\
    &  189 - 2060    &    Bates (1984) \\
O$_{2}$  &   198 - 546    &    Bates (1984) \\
Ar    &   140 - 2100   &    Bideau-Mehu et al. (1981) \\
CO$_{2}$ &  180 - 1700    &   Bideau-Mehu et al. (1973) \\
\cutinhead{King correction factor}
N$_{2}$    &  $\geq$ 200   &    Bates (1984) \\
O$_{2}$  & $\geq$ 200   &    Bates (1984) \\
Ar    &   all\tablenotemark{a}   &    Bates (1984) \\
CO$_{2}$ & 180 - 1700    &   Sneep \& Ubachs (2005) \\
\enddata
\tablenotetext{a}{$F_K = 1$ at all wavelengths}
\tablecomments{In unlisted spectral regions, quantities are assumed constant at the 
value of the closest wavelength for which data is available.}

\end{deluxetable}







\end{document}